\documentclass[12pt]{article}
\usepackage{amssymb,amsmath}
\usepackage{young}
\usepackage{graphicx}
\usepackage{ulem}
\usepackage{cite}
\begin{document} 
\begin{center} 
%\vspace{.1in} 
{\bf\Large {Connection between missing-charge in quasielastic electron scattering
and two(one) proton knockout reactions and the island-of-inversion} }
%{\bf\Large {Charge quenching of nuclei in quasi-elastic electron scattering} }
\end{center} 
\vspace{.1in} 
\begin{center} 
{\bf Syed Afsar Abbas}\\ 
Centre for Theoretical Physics, JMI University, New Delhi-110025, India\\
(email: drafsarabbas@gmail.com)\\
\vskip .1 cm
{\bf Usuf Rahaman, Anisul Ain Usmani}\\
Department of Physics, Aligarh Muslim University, Aligarh-202002, India\\

\end{center} 
%\vspace{.1in} 
\begin{center} 
{\bf Abstract} 
\end{center} 
%\vspace{.1in}
Our model, based on Quantum Chromodynamics, emphasizes the role of triton clustering in nuclei.
As there is good empirical support of triton clustering in nuclei, we may even treat this as a good
phenomenological working hypothesis as well. 
Here we show how our model finds remarkable success in being able to 
provide a consistent and interconnected understanding of such diverse empirical realities as the missing electric-charge
in quasielastic  electron scattering, and simultaneously, also the puzzling data within the studies of exotic nuclei. 
This gives us confidence, that
our model should be taking account  of the correct and proper degrees of freedom, to physically specify and explain, the total physical reality as manifested in the nuclear medium.

\newpage

Baryons, mesons, quarks and gluons, are normally taken as 
the relevant degrees of freedom for studies of the nucleus and the Quark Gluon Plasma (QGP).
However interestingly, clusters like alpha, well known to be prominent
in the studies of even-even nuclei, are being found to be significant even in QGP
as shown recently by Li, Zhang and Ma ~\cite{Ma}.
In this paper we continue our studies of the significance of the less well known, though actually very
fundamental, the triton cluster; in the nucleus.

In quasielastic electron scattering, both the momentum and the
energy transfers may be varied independently, within the constraints of the space-like four-momentum transfer. The quasielastic cross section may be separated into $R_L$, the longitudinal and $R_T$, the transverse response functions. As such
quasielastic electron scattering provides a
window into the field of nuclear medium effects including that of non-nucleonic
degrees of freedom in nuclei.
First, the SACLAY data indicated missing charge in quasielastic scattering ~\cite{Gerard}.
The status of quasielastic electron scattering on nuclei until 1993, was comprehensively summarized by Richter ~\cite{Richter}.
He discussed the longitudinal response data, which indicated puzzling charge quenching in nuclei, made apparent by the model independent Coulomb Sum Rule (QSR) analysis. Missing nuclear electric charge in the CSR ranged from $0\%$ in $^{3}He$,
\;to\;$40\%$ in $^{40}Ca \;and\;^{56}Fe$, \;and\;to about $50\%$  in $^{208}Pb$. 
Theoretical status of this puzzle in 1993, within the Quantum Hadrodynamics model may be gathered from the
illuminating review article by Wehrberger ~\cite{Wehrberger}.

By the beginning of the 21st century, several quasielastic experiments were carried out at BATES, SACLAY and SLAC; to extract 
$R_L \;and\;R_T$ for various nuclei. For medium and heavy nuclei, these experiments gave conflicting results, from 
full saturation of the CSR to $50\%$ quenching. Theoretical scenario was equally ambiguous.
Morgenstern and Meziani in 2001, came up with strong and convincing arguments~\cite{Morgenstern}, 
to show that the longitudinal response function $R_L$, for $^{40}Ca$, $^{48}Ca$, $^{56}Fe$, 
$^{197}Au$, $^{208}Pb$, $^{238}U$ are quenched, and that the CSR is not saturated.
However, at present, both the experimental and the theoretical situations, still remains ambiguous.
We expect that the Jefferson Laboratory experiments may provide
the final resolution of this puzzle. The JLAB experimental situation as of now, may be
gleaned through Paolone's recent contribution (iron,carbon)~\cite{Paolone},
and that of Lin (4He)~\cite{Lin}.

We ask whether the above quasielastic missing electric charge is also manifesting itself in some other nuclear physics scenario?
The so called island-of-inversion and the study of exotic nuclei offer possibilities.
The exotic nuclei have been the focus of intense interest
in nuclear physics, as these have opened up several puzzles in terms of halo nuclei, and new magic numbers, etc.
Within the context of new magic numbers, one-proton and two-proton knockout reactions have been playing useful
role. Fridmann {\it et al.} ~\cite{Fridman},~\cite{Fridman2}
 through studies of two-proton knockout reaction $^{44}_{16}S_{28} \rightarrow ^{42}_{14}Si_{28}$,
 presented a strong empirical evidence in support of magicity and sphericity of $^{42}_{14}Si_{28}$.
 What they essentially explored was the amazing persistence of
 the unique exotic nucleus $^{42}_{14}Si_{28}$, as a stable structure
 within the nucleus $^{48}_{20}Ca_{28}$; even after stripping off six-protons 
 through the isotonic chain:
 $^{48}_{20}Ca_{28} \rightarrow ^{46}_{18}Ar_{28}
 \rightarrow ^{44}_{16}S_{28} \rightarrow ^{42}_{14}Si_{28}$.
 Thus it is the novel stability of proton shell closure at
 Z=14 in $^{42}_{14}Si_{28}$, which is playing such a dominant role
 in ensuring its double magicity within $^{48}_{20}Ca_{28}$.
Thus Z=14 appears as a new magic number here.

In a similar situation within the sphere of the island-of-inversion, there is another two-proton knockout reaction
$^{32}_{12}Mg_{20} \rightarrow ^{30}_{10}Ne_{20}$ ~\cite{Fullon}.
The measured cross section was significantly suppressed compared to the theoretical calculation, indicating reduced overlap
between the initial and the final nucleus. Could this, and 
the above $^{42}_{14}Si_{28}$ case, be related to the missing/quenched charge of quasielastic scattering?

In this paper, we would like to present a combined analysis of the issue of missing charge and the above new magic numbers puzzle in 
exotic nuclei, within
our Quantum Chromodynamics based model, which has found successful applications in the nuclear medium.

Now everyone knows that $^{4}He$ is made of two protons and two neutrons, each sitting in $1s_{\frac{1}{2}}$ shell.
Its strong binding is believed to be the reason why light even-even nuclei may be considered to be alpha-cluster nuclei.
Hence it may come as surprise to many, that even as light a nucleus as  $^{4}He$,
has A=3 clusters sitting within it.
It has been convincingly demonstrated ~\cite{Kanada} 
that, contrary to expectations, the ground state of 
${^{4}_{2} He_{2}}$ contains none of above four nucleon configuration in the $1s_{\frac{1}{2}}$ orbital,
and very little of deuteron-deuteron
configuration either; and the same is actually built upon h-n and t-p ($ triton\;t \sim ^{3}_{1}H_{2}\;and\;
helion\;  h \sim ^{3}_{2}He_{1}$)
configurations ~\cite{Langanke}. 
This can be seen through Table below which lists just a few relevant configurations from the pioneering  work of Kanada, Kaneko and 
Tang ~\cite{Kanada}, where they stated, "The p+t and n+h cluster configurations,.... is the dominant component
in both the ground state and first excited sate $0^+$ of  $^{4}He$."
\vskip 1 cm

\begin{tabular}{c c c}
\hline\hline
Configuration                    &  Ground 0+ (MeV)  &  Excited 0+ (MeV) \\
\hline
   (dd)                          &     -19.21        &       -3.22       \\
(dd)+(d$^{*}$d)+(d$^{*}$d$^{*}$) &     -21.90        &       -3.68       \\
(pt)+(nh)                        &     -25.17        &       -5.10       \\
Full Calculation                 &     -26.50        &       -6.43       \\only 
Experiment                       &     -28.30        &       -8.30       \\
\hline\hline
\end{tabular}
\vskip 1 cm

Hence, if we consider only the ground state ${0_1}^+$ and the first excited state ${0_2}^+$
of ${^{4}_{2} He_{2}}$, then their wave function are naturally given as,

\begin{equation} 
{{\Psi}_{{0_1}^+}} =  
{{ [ {{\psi}_{n}}  \otimes {{\psi}_{h}} - {{\psi}_{p}} \otimes 
{{\psi}_{t}} ] } 
\over {\sqrt {2}}}
\end{equation}
\begin{equation}
{{\Psi}_{{0_2}^+}}  =
{{ [ {{\psi}_{n}}  \otimes {{\psi}_{h}} + {{\psi}_{p}} \otimes 
{{\psi}_{t}} ] }
\over {\sqrt {2}}}
\end{equation}

So when we say that $^{16}O$ is made up of four-$^{4}He$, or that $^{40}Ca$ is made up of ten-$^{4}He$,
we are not talking of $^{4}He$ being just two-protons and two-neutrons sitting in the orbital $1s_{\frac{1}{2}}$,
but have the configuration given in eqns. 1 and 2. As the two configurations given in eqns. 1 and 2,
are intrinsically correlated with each other, what happens to configuration in eqn. 2 in alpha-cluster nuclei
like $^{16}O$ and $^{40}Ca$? As such it would be impossible to ignore quantum mechanical interference effects of this 
${0_2}^+$ state in these several alpha-cluster nuclei. Alpha decay of heavy nuclei, as well known however, 
has to overcome only the coulomb barrier.
There is no evidence of the presence of the ${0_2}^+$ state in the escaping alpha-nuclei.
Thus, though the structure of alpha is as per eqn. 1 and 2, there does not appear to be any perceptible effect of the mixed state
of alpha in nuclear physics. This conclusion gains greater significance in the light of the recent work
in looking for alpha-clustering within $^{16}O$, in seeking a consistent understanding of the Quark Gluon Plasma~\cite{Ma}

Note that there shall not be any effect of ${0_2}^+$ state in alpha-clustering in  nuclei, either when it goes to very-high energy
or infinity, or when it collapses to the ground state. We ignore the first option as being highly unlikely.
The collapse of ${0_2}^+$ onto ${0_1}^+$, would mean that the wave functions given in eqn. 1 and 2, would become either
${{\Psi}_{{0_2}^+}} =  {{\Psi}_{{0_1}^+}} \sim
{{\psi}_{p}}  \otimes {{\psi}_{t}}\;\;OR\;\; \sim {{\psi}_{n}}  \otimes {{\psi}_{h}}$.
The Coulomb energy (self-energy) would favour the proton-triton composite structure. Hence we propose
that alpha in nuclear medium  would exhibit a propensity of being a composite of proton and triton. Thus alpha plus two-neutrons would behave as a composite of two tritons;
\begin{equation}
^{4}He \sim {{\psi}_{p}}  \otimes {{\psi}_{t}} = (p-t)\;;\;\;\;\;\;^{4}He + 2n \sim {{\psi}_{t}}  \otimes {{\psi}_{t}}\;= (t-t)
\end{equation}
Thus triton clustering under appropriate conditions, should be a common feature in a nucleus.
We may treat the structure of alpha given in eqn. 3,
as a new phenomenological feature of the nuclear medium. 
Thus at one working level one accepts triton clustering in nuclei, and apply this idea in the nucleus, and see
how good it works.
However, we can do better, as we have actual theoretical justification of this picture, which we discuss below.

This new phenomenological principle, however, was earlier introduced by the author SAA, as a result arising from
arguments based on hidden colour and Quantum Chromodynamics~\cite{Abbas2001}.
Its has proven its usefulness by explaining various nuclear reality as of halo structures, new magic  numbers, 
nuclear fusion etc.~\cite{Abbas2001},~\cite{Abbas2004},~\cite{Abbas2005},
~\cite{Abbas2011},~\cite{Abbas2018},~\cite{Abbas2019},~\cite{Abbas2020},~\cite{Abbas2016}.
The Relativistic Mean Field Theory has justified this model based conclusions ~\cite{Abbas2018},~\cite{Abbas2019},
~\cite{Abbas2020}. 

Hence let us  treat all $ {\rm ^{3Z}_{\:\:Z} X_{2Z}} $ 
nuclei as being a bound state of Z-number of
tritons (${^{3}_{1} H_{2}}$). 
Viewed in this manner, the relevant degrees of freedom are tritons which are treated as
"elementary" entities.
In analogy with the fact that we know as per mean field concept, 
that  a bunch of protons and neutrons in a nucleus, would 
create an average  binding potential for each nucleon, we assume 
that a bunch of tritons in a nucleus 
too would create an average binding potential for each triton in it.
Such a potential would bind tritons in these neutron rich nuclei with
${\rm ^{3Z}_{\:\:Z} X_{2Z}} = Z{\rm ^3_1H_2}$.
We extract one-triton separation energies of these pure triton constituent nuclei.

\begin{equation}
S_{1t}={\rm B}({\rm ^A_ZX_{2Z}})
-{\rm B}({\rm _{Z-1}^{A-3} Y_{2Z-2}})-{\rm B}(_1^3{\rm H}_2)
\end{equation}
where, ${\rm B}({\rm ^A_ZX_N})$ is the binding energy of the nucleus
${\rm ^A_ZX_N}$. 
Recently we have conducted a comprehensive theoretical study within 
the ambit of the field of the RMF model structure with three
good and successful interactions. We predicted ~\cite{Abbas2018} six prominent magic nuclei:
$_{\:\:8}^{24}{\rm O}_{16}$, $_{20}^{60}{\rm Ca}_{40}$,
$_{\:\:35}^{105}{\rm Br}_{70}$, $_{\:\:41}^{123}{\rm Nb}_{82}$,
$_{\:\:63}^{189}{\rm Eu}_{126}$ and $_{\:\:92}^{276}{\rm U}_{184}$. 

The experimental binding energies are not available beyond 
${\rm N}_t=17$ bound systems. 
Our prediction of double magicity of nucleus 
$ ^{24}_{8}O_{16}$, has been shown to hold good by Kanungo {\it et al.} ~\cite{Kanungo}.
We also made unique prediction, 
of magicity for $ ^{60}_{20}Ca_{40}$,
and which has since then been confirmed by Tarasov {\it et al.} ~\cite{Tarasov}.
These thus gave strong empirical support to our model and as shown in Fig. 1

Let us focus on $^{42}_{14}Si_{28}$ and $^{30}_{10}Ne_{20}$.
However, on closer scrutiny of the structure between the two extremes of 
the strongly magical nuclei: $^{24}_{8}O_{16}$  and $^{60}_{20}Ca_{40}$, 
we notice a prominent broad hump or "plateau of stability". 
This plateau of magicity is being 
defined by the two boundary towering peaks of magicity at $N_t = 8 \; = \;^{24}_{8}O_{16}$ 
and $N_t=20 \;=\;^{60}_{20}Ca_{40}$ respectively.
However equally significant, in defining this plateau of magicity, 
are the two boundary nuclei manifesting themselves 
as extremely-deep-trenches at $N_t = 9 \; = \;^{27}_{9}F_{18}$  
and $N_t = 21\; = \; ^{63}_{21}Sc_{42}$.
Thus both the boundary states, of the two towering peaks, and of 
the two deep trenches, provide a physically identifiable character
of magicity and stability to the whole range of nuclei, Z = 10, 12, 14, 16 and 18.

\begin{figure} 
\begin{center}
\vspace{0.5cm}
\includegraphics[scale=0.4]{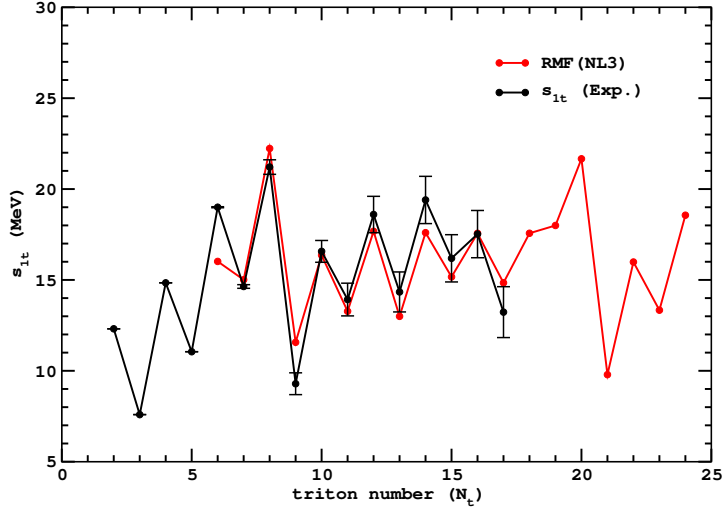}
\caption{  Triton separation energy in our triton-cluster model ~\cite{Abbas2018}  }
\label{Fig. 1}
 \end{center}
\end{figure}
 
Within our "plateau of magicity", in Fig. 1, there appears,
a slight kink ("slight", in the context of the large towering 
peaks at $^{24}_{8}O_{16}$  and $^{60}_{20}Ca_{40}$),
at $^{42}_{14}Si_{28}$, and which is somewhat more stable than the nuclei surrounding it, viz
$^{36}_{12}Mg_{24}$  and $^{48}_{16}S_{32}$. 
Thus $^{42}_{14}Si_{28}$ is even more magical than these two.

 Now, theoretically, how do we understand the extra stability and magicity of $^{42}_{14}Si_{28}$
 as exhibited in the experimental results of Fridmann et al.~\cite{Fridman},~\cite{Fridman2}?
 It has been a long standing paradigm in nuclear physics, that the central potential is proportional to the
 ground state baryon density, and the spin-orbit potential is proportional to the derivative
 of the same central potential. 
In a pioneering work, Todd-Rudel, Piekarewicz and Cottle  ~\cite{Todd-Rutel}, found that 
 the dramatic decrease in spin-orbit splitting as seen in exotic nuclei, is not caused by the neutron density in the nuclear 
 surface, but by proton density in the nuclear interior. With NL3 interaction, they ~\cite{Todd-Rutel} found
that as two-protons are removed from $^{48}Ca \rightarrow ^{46}{S}$, 
 the density of 
 $^{48}Ca$ transforms into a bubble/hole-like nucleus for 
 $^{46}{S}$ itself. But this fails to reproduce
 the amazing persistence of the nucleus $^{42}_{14}Si_{28}$ as a stable structure
 within the nucleus $^{48}_{20}Ca_{28}$.
 
 \begin{figure} 
  \vspace{0.5cm}
  \includegraphics[scale=0.4]{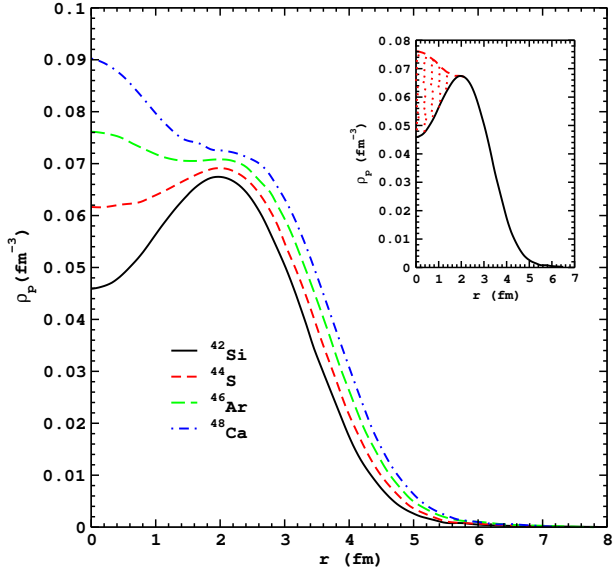}
  \caption{ Schematic point proton density plot of N=28 isotonic chain  $^{48}_{20}Ca_{28} \rightarrow ^{46}_{18}Ar_{28} \rightarrow ^{44}_{16}S_{28} \rightarrow ^{42}_{14}Si_{28}$ ~\cite{Piekarewicz}. Inset shows the schematic structure $^{48}_{20}Ca_{28} 
  \sim ^{42}_{14}Si_{28} + (6-p)$}
  \label{Fig. 3}
  \end{figure}

 Piekarewicz realized ~\cite{Piekarewicz}
 that the problem had to do with the fact that the NL3 interaction was  failing to produce the 
 $1 {d}^{\frac{3}{2}} - 2 {s}^{\frac{1}{2}}$ proton gap in $^{40}Ca$. In his remarkable paper ~\cite{Piekarewicz},
 he tweaked the NL3 parameters slightly, 
 in a minimal manner, so that this
 basic problem of the Calcium-chain was rectified.
 Thus he got consistent point proton density distribution of all the nuclei in the basic
 six-protons stripping isotonic chain:
 $^{48}_{20}Ca_{28} \rightarrow ^{46}_{18}Ar_{28} \rightarrow ^{44}_{16}S_{28} \rightarrow ^{42}_{14}Si_{28}$.
 His Fig. 4  ~\cite{Piekarewicz}, is reproduced here in our Fig. 1. In the inset we show how physically, this is equivalent
 to stripping six-protons from the stable core of $^{48}_{20}Ca_{28}$ itself.

 Most significant is that the density distribution of $^{42}_{14}Si_{28}$ has a hole at the centre.
 So it looks like a tennis-ball (bubble) like nucleus. This is a most direct confirmation of SAA's original predictions
 of 2001~\cite{Abbas2001}  . This confirms our above discussion of the extra stability and sphericity
 of $^{42}_{14}Si_{28}$ ~\cite{Abbas2020}.
Thus Z=14 is a new magic number, not because of 14-protons, but because of 14-tritons in 
triton-shell-model and with individual protons hidden inside each triton. 
Remember, that while proton size is $\sim 0.8 \; fm$, that of triton is $\sim 1.8\;fm$.

Thus $^{48}_{20}Ca_{28} \sim ^{42}_{14}Si_{28} + (6-p)$ in our model. Just like in $^{17}_{8}O_{8}$, we take a single neutron sitting outside a  closed shell of $^{16}_{8}O_{8}$ in shell model, here we take 6-protons sitting outside a strongly bound  nucleus of 
$^{42}_{14}Si_{28}$.
Thus in quasielastic scattering on $^{48}_{20}Ca_{28}$ we will see only these free 6-protons, and not the 20-protons as in the shell model
charge structure for 48-Ca. This is a clear and unique prediction of our model. We are not saying that this is the structure of 48-Ca in the ground state. But this is how it will look like at the internal energies relevant in the quasielastic scattering.

Next, in the two-proton knockout reaction
$^{32}_{12}Mg_{20} \rightarrow ^{30}_{10}Ne_{20}$ ~\cite{Fullon}.
Clearly in our picture ${^{30}_{10} Ne_{20}} = 10 \;{^{3}_{1} H_{2}}$ which as per Fig. 1,
is extra stable and thus magical. 
The giant peak at $N_t = 8 \; = \;^{24}_{8}O_{16}$  and the equally deep trench at $N_t = 9 \; = \;^{27}_{9}F_{18}$,
would help in enhancing the stability and magicity of $N_t = 10 \; = \;^{30}_{10}Ne_{20}$.
Therefore, given standard shell model wave function for the initial state, there would be a  reduced overlap
between the initial and the final nucleus. This explains why the  measured cross section was so much suppressed compared to the calculation.

Thus in quasielastic scattering on $^{40}_{20}Ca_{20}$, under excitations available in the scattering process, we expect
$^{40}_{20}Ca_{20} = ^{30}_{10}Ne_{20} + (10-p)$. Thus quasielastic scattering will only count ten-protons -
actually half of the expected twenty-protons. So we see wherefrom comes the electric charge quenching.

Now clearly also in our triton-clustering model
$^{56}_{28}Ni_{28} \rightarrow ^{42}_{14}Si_{28} + (14-p)$. Thus we can understand as to how half the charge shall only be available 
in quasielastic electron scattering on $^{56}_{26}Fe_{30}$.

Thus our model is able to explain the quenching in medium heavy nuclei. For heavy nuclei, clearly with appropriate modelling within our 
triton-clustering picture, we may reproduce missing charges. Also note that
of course there 
shall be no missing charge in $^{3}_{1}He_{2}$, as actually observed in experiments
~\cite{Gerard},~\cite{Richter},~\cite{Wehrberger}.
The extra stability of ${^{30}_{10} Ne_{20}} = 10 \;{^{3}_{1} H_{2}}$ and those of other nuclei as displayed in Fig. 1, would enable us to understand the basic cause of the existence of
the island-of-stability.

Our model of triton clustering in nuclei, initially arose from a proper formulation of the hidden colour concept and
Quantum Chromodynamics ~\cite{Abbas2001}. We have shown that triton clustering is empirically well justified too ~\cite{Abbas2011}.
Empirical justification of triton clustering also means that we can treat it as a suitable phenomenological principle.
Thus one may take eqn. 3 as a successful phenomenological concept, which is justified due to its physically successful applications
in exotic nuclei. Here we have shown how the same picture finds successful application in quasielastic electron scattering.

It is remarkable that in our model, we are able to 
provide a consistent and interconnected understanding of such diverse empirical realities as the quasielastic 
electron scattering and the puzzling data within the studies of exotic nuclei. Are nucleons, pions/mesons, quarks and gluons,
the only basic degrees of freedom relevant in nuclei at low energies or in the Quark Gluon Plasma (QGP)?
Some recent work within QGP by Li, Zhang and Ma~\cite{Ma} however, requires alpha clusters in QGP.
Above, we have shown that tritons too, should be relevant under appropriate conditions, in a nucleus.
This gives us confidence, that
our model should be taking account  of the correct and proper degrees of freedom, to physically specify and explain, the total physical reality as manifested in the nuclear medium.
 
%\newpage

\end{document}